\documentclass[revtex,showkeys,preprintnumbers,amsmath,amssymb,nofootinbib]{revtex4}

\usepackage[utf8]{inputenc}  
\usepackage{graphicx}
\usepackage{dcolumn}
\usepackage{bm}

\usepackage{amsmath}
\usepackage{bbm}             
\usepackage{bm}              
\usepackage{color}
\usepackage{latexsym,amsfonts}
\usepackage{graphicx}
\usepackage[                 
	pdfsubject={Vorlesungsskriptum},
	pdfauthor={Manfried Faber},
	pdfkeywords={meta, information, pdf, hyperref, latex},
        colorlinks=true,
	urlcolor=blue,
	linkcolor=Red2,
        citecolor = magenta,
        pdfpagelabels,
        pdfstartview = FitH,
        bookmarksopen = true,
        bookmarksnumbered = true,
        plainpages = true,
        hypertexnames = false]{hyperref}

\providecommand{\href}[2]{#2}   
\definecolor{Blue2}{rgb}{0.,0.,0.8125}
\definecolor{Brown3}{rgb}{0.625,0.25,0.}
\definecolor{Cyan4}{rgb}{0.,0.56,0.56}
\definecolor{Green4}{rgb}{0.,0.56,0.}
\definecolor{LtBlue}{rgb}{0.27,0.42,0.52}
\definecolor{Magenta4}{rgb}{0.5625,0.,0.5625}
\definecolor{Red2}{rgb}{0.8125,0.,0.}

\DeclareUnicodeCharacter{2212}{-}

\newcommand{\ist}[1]{\overset{\footnotesize(\ref{#1})}{=}}
\newcommand{\iist}[2]{\overset{^{(\ref{#1})}}{\underset{^{(\ref{#2})}}{=}}}

\begin{document}

\title{Stationary Schrödinger Equation and Darwin Term from Maximal Entropy Random Walk}

\author{Manfried Faber}
\affiliation{Atominstitut, Technische Universit\"at Wien,
Stadionallee 2, 1020 Wien, Austria}

\date{\today}

\begin{abstract}
We describe particles in a potential by a special diffusion process, the maximal entropy random walk (MERW) on a lattice. Since MERW originates in a variational problem, it shares the linear algebra of Hilbert spaces with quantum mechanics. The Born rule appears from measurements between equilibrium states in the past and the same equilibrium states in the future. Introducing potentials by the observation that time, in a gravitational field running in different heights with a different speed, MERW respects the rule that all trajectories of the same duration are counted with equal probability. In this way, MERW allows us to derive the Schrödinger equation for a particle in a potential and the Darwin term of the nonrelativistic expansion of the Dirac equation. Finally, we discuss why quantum mechanics cannot be simply a result of MERW, but, due to the many analogies, MERW may pave the way for further understanding.
\end{abstract}

\maketitle
\section{Introduction}\label{sec:Introduction}
The phenomena at atomic and subatomic scales are very successfully described by quantum mechanics and the Schrödinger equation~\cite{griffiths2018}. Despite existing now for almost 100~years, their interpretation is still a topic of intensive discussion~\cite{EmergingQuantum2014,cook2018,ralston2018,DongShi-Hai2007}. We know how to use the equations and to make predictions, but there is no agreement yet on how nature produces the effects described by quantum mechanics. Since we have no satisfactory answer to this question, we have to collect the available arguments for a future consistent picture which explains the appearance of quantum phenomena.

Most of the community is satisfied with wave-particle duality, based on the experience that quantum phenomena can be explained by waves, but detection at low intensities is registered locally as being produced by particles. The question is raised whether this is related to the atomic structure of detectors. On the other hand, some authors are inclined to the explanation by wave phenomena. They derive the Schrödinger equation from the wave equation as an approximation for potentials much smaller than the rest mass, see e.g\@.,~Section~6.5 of Ref.~\cite{ralston2018}. They explain the Born rule by the quadratic Hamiltonian for electromagnetism and the linear superposition principle from the linear equations, which are valid for waves.

Waves and particle approaches do not take into account that there are phenomena between waves and particles, as well as topological solitons, which are formulated as field models. By their topological properties, special waves characterised by topological quantum numbers become spatially concentrated. These properties of topological solitons may explain why we observe both wave and particle properties. An interesting experiment which demonstrates how wave and particle properties may cooperate is the oil drop experiment invented by Couder~\cite{couder2005}. A numerical simulation of travelling waves, solitons, and radiating excess energy as dispersive waves is described in Sect.~VI.5 of the book by Komech~\cite{Komech2022}.

In this article, I intend to offer another insight into quantum mechanical processes by employing the maximal entropy random walk (MERW), which is one of the approaches in studying diffusion processes in complex systems \cite{Burda2009,Sinatra2011,Ochab2012,ochabburda2013}. It models processes where the stationary probability of finding a particle localises in the largest nearly spherical region that is free of defects. Thus, MERW differs on irregular lattices from the generic random walk (GRW), where, at each time step, particles go with equal probabilities to any one of its nearest neighbours. This allows us to derive the stationary Schrödinger equation, to explain the Born rule \cite{duda2008,duda2021diffusion,duda2023fourdimensional} and, as shown in this article for the first time, to obtain the Darwin term, which usually appears in a relativistic formulation only \cite{Darwin1928,sakurai2006}. The rules of linear algebra follow as a consequence of the solution of a variational problem.

MERW is a Markov chain, its time evolution of trajectories depends on the present state only and not on the past. As discussed in Section \ref{SecMERW}, it is defined by a step matrix $M$, acting on vectors of position distributions, known as amplitudes. 
In distinction from other random walks, it is also defined by the additional request that all paths taking the same time are assumed to have the same probability. Such a uniform distribution of trajectories has a maximal entropy and justifies the choice of the name~\cite{Burda2009}. For particles in a potential, the solutions of the variational problem are given by the eigenfunctions of $M$. They provide the ensemble of paths and its equilibrium distribution. Measurements are conducted between the equilibrium distributions in the past and future. From the identity of these distributions follows the time symmetry of the ensemble of trajectories forward and backward in time and the Born rule. In Section \ref{sec3}, we apply MERW to the free motion in a box and to motion in a potential. As long as the free motion is not inhibited by boundaries or a potential  and if the diffusion constant is determined under the assumption that in reduced Compton, time particles may move a distance of a reduced Compton wavelength, we find that the amplitudes obey the time dependent Schrödinger equation continued in 
 imaginary time. In the case of the influence of a potential, we find that the infinite time limit, the stationary distribution, corresponds to the solution of the time-independent Schrödinger equation. Expanding the solution of the variational problem up to $\alpha_f^4$, we obtain the Darwin term. In Section \ref{sec:Excited states}, we discuss the problems with the probability interpretation of excited states and their technical solution.

We formulate MERW on a cubic homogeneous lattice with spacing $\delta$. Expanding to the second order in $\delta$, we can compare the terms with the three terms which we obtain in the nonrelativistic expansion of the Dirac equation. Despite the nonrelativistic formulation, MERW can describe Zitterbewegung and obtain the correct form of the Darwin term. From the nonrelativistic treatment, it is understandable that the other two terms differ.

\section{Method}\label{SecMERW}
In Refs.~\cite{duda2008,Burda2009}, MERW was introduced as a stochastic process in discrete space and time on general graphs, which are structures made of vertices (nodes) $x$ and edges $xy$. In a single time step of a Markov chain, the particle may hop along an edge from $x$ to a neighbouring node $y$ with a probability described by the stochastic (transition; Markov) matrix $S_{xy}$ independent of the past history, leaving the probability distribution $\rho_x$ invariant; see Equation~(\ref{SInvar}). Equal probabilities $S_{xy}$ outgoing from a vortex $x$ to all neighbours $y$ define the commonly used generic random walk (GRW). Due to this equality, GRW maximises the entropy locally, whereas we will define MERW such that it maximises the entropy of the ensemble of paths of an equal length.

Starting with an initial distribution $|i\rangle$ of positions MERW counts with the amplitudes
\begin{equation}\label{WellenFunkt}
\psi_i(x):=\langle x|i\rangle
\end{equation}
the number of paths starting at $x$. MERW is based on a step matrix $M$ with non-negative matrix elements $\hat M(y,x):=\langle y|M|x\rangle$ which is defined by
\begin{equation}\label{kSchritte}
|f\rangle=M^k|i\rangle\quad\Leftrightarrow\quad
\psi_f(y)=\sum_x\hat M^k(y,x)\psi_i(x),
\end{equation}
the number of paths ending after $k$ time steps at $|y\rangle$. The elements of $M$ are defined in Equation~(\ref{simpleMatrix}) for particles in a box and in Equation~(\ref{MExp}) for particles in a potential $V$.

Constructing the possible paths with this step matrix $M$ guarantees that all trajectories which take the same time have an equal probability. This uniform distribution of paths has a maximal entropy and justifies calling the method ``Maximal Entropy Random Walk''. The principle of maximum entropy~\cite{Jaynes1957} states that the probability distribution which best represents the current state of knowledge about a system is the one with the largest entropy. For time-reversal invariant processes, we can assume that steps away and back are equally likely and thus that the matrix $M$ is symmetric.

Our interest concentrates on the stationary states of this diffusion process. We obtain them from the eigenvalue equation of the symmetric step matrix $M$
\begin{equation}\label{EGlM}
M|i\rangle=\lambda_i|i\rangle\quad\Leftrightarrow\quad
\underbrace{\langle x|M|i\rangle}_{\hat M\psi_i(x)}
=\lambda_i\underbrace{\langle x|i\rangle}_{\psi_i(x)}.
\end{equation}

For any finite number of time steps, these states are normalisable and define orthogonal rays in a real vector space

The set of orthonormal eigenvectors $|i\rangle$
\begin{equation}\label{Orthogon}
\langle i|j\rangle=\delta_{ij}
\end{equation}
defines a complete basis
\begin{equation}\label{iVollst}
\sum_{i=1}^N|i\rangle\langle i|=\mathbbm 1_n.
\end{equation}

If the limit for an infinite number of steps corresponds to a bound state, this limit is also normalisable.

Due to the request of the highest entropy of the ensemble of paths, the state $|0\rangle$ with the highest eigenvalue $\lambda_0$ is stable,
\vspace{-9pt}

\begin{equation}\begin{aligned}\label{stable}
\langle x|0\rangle\langle 0|y\rangle&=\lim_{k\to\infty}\sum_i
\langle x|\underbrace{\left(\frac{\lambda_i}{\lambda_0}\right)^k}_{\to\delta_{i0}}
|i\rangle\langle i|y\rangle
\ist{EGlM}\lim_{k\to\infty}\langle x|\left(\frac{M}{\lambda_0}\right)^k
\sum_i|i\rangle\langle i|y\rangle=\\&\ist{iVollst}
\lim_{k\to\infty}\langle x|\left(\frac{M}{\lambda_0}\right)^k|y\rangle,
\end{aligned}\end{equation}
the other eigenstates are metastable. The Perron–Frobenius theorem in the matrix theory states for which matrices there exists a highest real eigenvalue $\lambda_0$ with non-negative components $\langle x|0\rangle\ge0$, i.e., without nodes.

Due to the iterability
\begin{equation}\label{BezGleichGewicht}
\langle x|0\rangle\ist{EGlM}
\langle x|\underbrace{\frac{M}{\lambda_0}\sum_y|y\rangle\langle y|}_{1}0\rangle
\ist{EGlM}\langle x|\underbrace{\frac{M}{\lambda_0}\sum_y|y\rangle\langle y|}_{1}
\underbrace{\frac{M}{\lambda_0}\sum_z|z\rangle\langle z|}_{1}0\rangle
\end{equation}
we are able to find the stochastic matrix $S_{xy}$.~Stochastic matrices  are also called transition or Markov matrices and determine the transition probabilities of Markov chains. As the matrix elements are probabilities, their values satisfy\begin{equation}\label{SElem}
0\le S_{xy}\le1.
\end{equation}

According to Equation~(\ref{SGibtEins}), the matrices $S$ are leftstochastic, this means that their columns form probability vectors and thus add up to $1$; see Equation~(\ref{SGibtEins}). The $1$-norm is $||S||_1:=\max\sum_y|S_{xy}|=1$. Since the spectral radius of a matrix is smaller than the norm, all eigenvalues of $S$ are smaller or equal to $1$, which leaves a probability distribution $\rho_x$ invariant 
\begin{equation}\label{SInvar}
\rho_z=\sum_x\rho_xS_{xz}=\sum_x\rho_x\sum_yS_{xy}S_{yz}
\end{equation}
and with a sum over final states of value $1$ from a given initial site $x$\begin{equation}\label{SGibtEins}
\sum_yS_{xy}=1.
\end{equation}

Equations~(\ref{SInvar}) and (\ref{SGibtEins}) can be fulfilled by
\begin{equation}\label{pDef}
\rho_x:=\frac{\langle0|x\rangle\langle x|0\rangle}{\langle0|0\rangle},\quad S_{xy}:=
\frac{1}{\langle x|0\rangle}\langle x|\frac{M}{\lambda_0}|y\rangle\langle y|0\rangle,
\end{equation}
where the denominators normalise the number of initial positions in the ground state to 1. We obtain, for the iterated stochastic matrix
\begin{equation}\label{SIterNorm}
\sum_y(S^k)_{xy}\ist{pDef}1
\end{equation}
and the condition
\begin{equation}\label{rhoNorm}
\sum_x\rho_x\ist{pDef}\frac{\langle 0|\overbrace{\sum_x|x\rangle\langle x}
^{\ist{iVollst}1}|0\rangle}{\langle0|0\rangle}\ist{iVollst}1
\end{equation}
which is requested for a probability distribution.

The entropy growth in MERW processes is explicitly defined in Appendix~\ref{SecEntropy}. There are also examples to show that the entropy is maximal for the states with the highest eigenvalue $\lambda_0$.

An interesting discussion about the interpretation of the wave function in quantum mechanics, as the probability amplitude that the system is in a certain configuration, can be found in the works of Mara Beller~\cite{Beller1999}.

\section{Applications of MERW}\label{sec3}
\subsection{Free Motion in a Box}
The first impression about the method that we have for free motion is in a box of size $L^D$ and infinitely high walls in $D$ dimensions. We are using the sites $\vec x$, coordinates $x_d$, unit vectors $\hat d$ and lattice spacing $\delta$ for $(N+1)^D$ sites
\begin{equation}\label{Gitterkoord}
x_n^d=\left(n^d-\frac{N}{2}\right)\delta,\quad n^d\in\{0,1,2\dots N\},\quad
\delta=\frac{L}{N}.
\end{equation}

We describe free motion with a step matrix $M$, e.g., by
\begin{equation}\label{simpleMatrix}
\langle\vec x|M|\vec y\rangle=\langle\vec y|M|\vec x\rangle=
\begin{cases}
1\quad\textrm{for}\quad |\vec x-\vec y|=\delta,\quad
\vec x\textrm{ and  }\vec y\textrm{ not at the boundary},\\
0\quad\textrm{else}.
\end{cases}
\end{equation}

We can start the time evolution with some small initial distribution around $\vec\mu$, somewhere in the middle of the lattice, e.g.,
\begin{equation}\label{AnfPsi}
\mathring\psi(\vec x):=\prod_{d=1}^D\delta_{x_d,\mu_d},
\end{equation}
and evolve the amplitudes by
\begin{equation}\label{Zeitent}
  \psi(\vec x,\tau):=
  \sum_{\vec y}\hat M^\tau(\vec x, \vec y)\,\mathring\psi(\vec y).
\end{equation}

According to the central limit theorem, the amplitudes tend asymptotically to the Gaussian distributions
\begin{equation}\label{psiGauss}
\phi(\vec x,\tau)\propto\exp\{-\frac{(\vec x-\vec\mu)^2}{2\sigma_\tau^2}\},
\end{equation}
as long as the distribution does not reach the boundaries.

Due to the additivity of variances, the variance of a coordinate increases in every time step by the same amount $\delta\sigma^2$ leading after $\tau$ time steps to the variance
\begin{equation}\label{sigmatau}
\sigma_\tau^2\ist{sigma1}\tau\,\delta\sigma^2.
\end{equation}

For the step matrix $M$ of Equation~(\ref{simpleMatrix}), we obtain
\begin{equation}\label{sigma1}
\delta\sigma^2\ist{simpleMatrix}\frac{2\delta^2}{2D}=\frac{\delta^2}{D}.
\end{equation}

Normal diffusion, obeying Fick's law, would approach a constant distribution, whereas MERW ends in the state of the highest entropy, with an amplitude proportional to a product of cosine functions
\begin{equation}\label{LoesFreiSchroe}
\psi_0(x)\propto\prod_{d=1}^D\cos(k_0x^d)\quad\textrm{with}\quad k_0
=\frac{\pi}{L}.
\end{equation}

\begin{figure}[h!]
\includegraphics[scale=0.5]{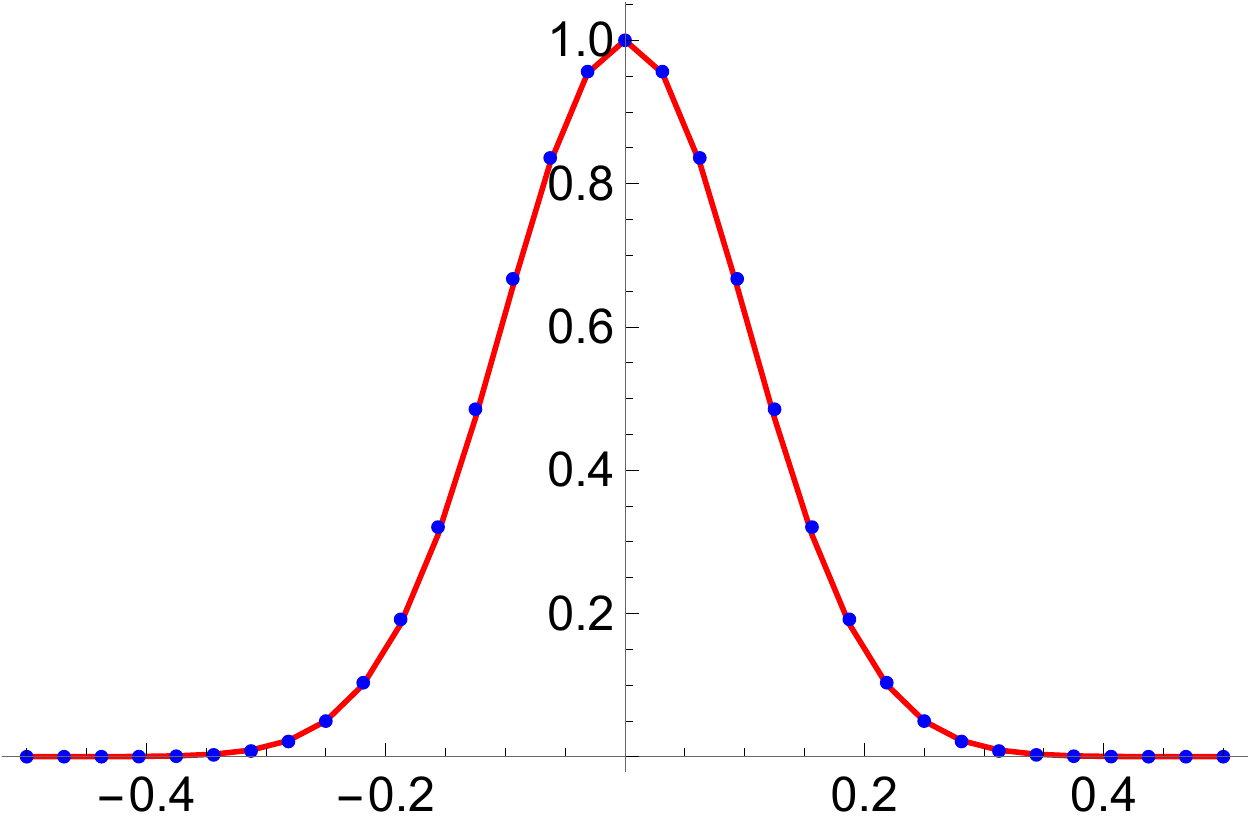}\hspace{10mm}
\includegraphics[scale=0.5]{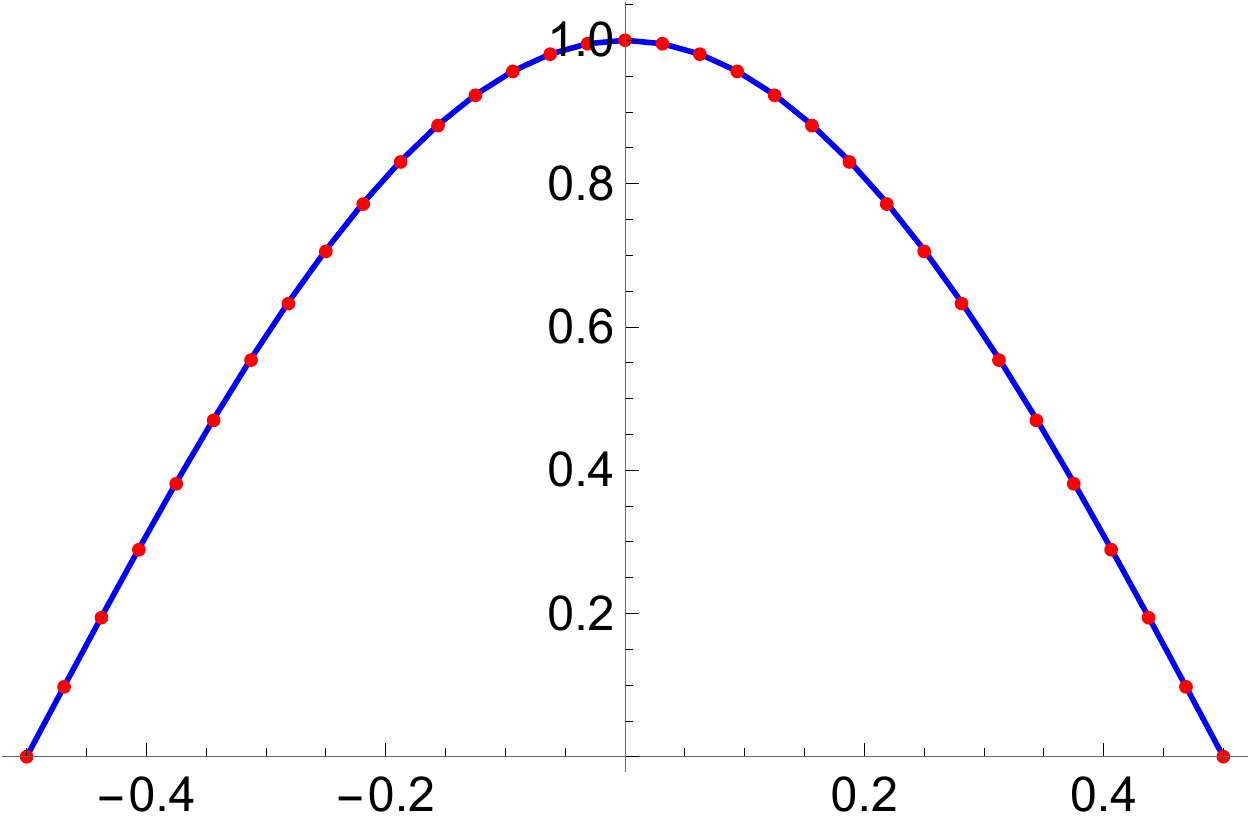}
\caption{(\textbf{Left}): The blue dots depict the amplitudes $\psi(x,16)/\psi(0,16)$ according to Equation~(\ref{Zeitent}) and the red line joins the dots of $\phi_{16}(x)/\phi_{16}(0)$ according to Equation~(\ref{psiGauss}). (\textbf{Right}): After 256 time steps $\psi(x,256)/\psi(0,256)$ (red dots), the cos-function approaches $\cos(\pi x)$, as shown by a \mbox{blue line}.}\label{psirand} 
\end{figure}

To check this behaviour, we have evolved, in Figure~\ref{psirand}, amplitudes in one dimension for $L=1$, $N=32$ and $\mu=0$, according to Equation~(\ref{Zeitent}). In the left diagram, we compare the amplitudes $\psi(x,16)/\psi(0,16)$ (blue dots), after 16 time steps, before the distribution arrives at the boundary, with the corresponding Gaussian $\phi_{16}(x)/\phi_{16}(0)$ of free diffusion (red line). We find a similar nice agreement in the right diagram after 256 time steps between $\psi(x,256)/\psi(0,256)$, with red dots, and the quantum mechanical ground state $\cos(\pi x)$, the blue line, in a box of size 1. According to Equation~(\ref{simpleMatrix}), there are no paths leading into the walls. This is why we observe in the diagram on the right that particles cannot accumulate near the walls where the freedom of movement is restricted. There are many more paths that lead away from the walls, into areas where particles can move freely.
\vspace{-4PT}

The normalised time-dependent Gaussian
\begin{equation}\label{normGauss}
\phi(\vec x,\tau):=\frac{1}{(2\pi\,\tau\,\delta\sigma^2)^D}
\exp\{-\frac{\vec x^2}{2\tau\,\delta\sigma^2}\}
\end{equation}
is a solution of the diffusion equation
\begin{equation}\label{DiffusionsGl}
\partial_\tau\phi(x,\tau)=D_d\,\Delta\,\phi(x,\tau)\quad\textrm{with}\quad
D_d=\frac{\delta\sigma^2}{2}
\end{equation}
with the diffusion constant $D_d$. Introducing the imaginary time $\tau=\mathrm i\frac{t}{\delta t}$ and the time step $\delta t$
\begin{equation}\label{imDiff}
\mathrm i\partial_t\phi(x,t)=-\frac{D_d}{\delta t}\,\Delta\,\phi(x,t).
\end{equation}
the diffusion equation agrees with the time-dependent Schrödinger equation for a free particle~\cite{Schrodinger:1931aa}, if we choose
\begin{equation}\label{DiffQM}
D_d:=\frac{\hbar}{2m}\,\delta t
\end{equation}
leading to
\begin{equation}\label{Hdm}
\frac{\hbar}{m}\iist{DiffQM}{DiffusionsGl}\frac{\delta\sigma^2}{\delta t}.
\end{equation}

This fraction can be expressed by the ratio of the square of the reduced Compton wave length $\bar\lambda_c$ and the reduced Compton time $\bar t_c$
\begin{equation}\label{ComptVerh}
\frac{\hbar}{m}=\frac{\bar\lambda_c^2}{\bar t_c}\quad\textrm{with}\quad
\bar\lambda_c:=\frac{\hbar}{mc},\;\bar t_c:=\frac{\hbar}{mc^2}
\end{equation}

Inserting Equation~(\ref{Hdm}), we conclude
\begin{equation}\label{CompVerh}
\frac{\delta t}{\bar t_c}\iist{Hdm}{ComptVerh}\frac{\delta\sigma^2}{\bar\lambda_c^2}.
\end{equation}

In the $n^2$ reduced Compton times, the diffusion therefore propagates in every direction $n$ and reduced Compton wave lengths; the single step is conducted with velocity $c_0/n$. What is appropriate for quantum physics could be $n=1/\alpha_f=137.036$, where $\alpha_f$ is Sommerfeld's fine structure constant. Since the further results are not influenced by the factor $n$, we choose further results 
 for simplicity $n=1$, and obtain
\begin{equation}\label{deltatdelta}
\delta t=\bar t_c=\frac{\hbar}{mc^2},\quad \delta\sigma^2=\bar\lambda_c^2
\ist{ComptVerh}\frac{\hbar^2}{m^2c^2}.
\end{equation}

After reaching the stationary solution, depicted in the right diagram of Figure~\ref{psirand}, the time derivative in the diffusion Equation~(\ref{DiffusionsGl}) vanishes, and the equation agrees with the time-independent Schrödinger equation. The solution is now influenced by boundary conditions. The problem has turned to a boundary value problem, characterised by eigenvalues and eigenfunctions, and by rays in a vector space.

\subsection{Motion in a Potential}
From free motion, we had the idea that the solutions of MERW may agree with solutions of the stationary Schrödinger equation. In this section, we will investigate under which conditions we can obtain from MERW the stationary Schrödinger equation for a particle in \mbox{a potential.}

In classical and quantum mechanics, we use time as a universal parameter, despite knowing that time in different gravitational potentials runs with a different speed~\cite{chou24092010}. With the concept of energy as a factor multiplying this universal time~\cite{FaberZeit}, we describe moving particles measuring the action $S=\int\mathrm dt L$ in units of $\hbar$. In a potential $V$ and for vanishing kinetic energy $T$, the Lagrangian reads
\begin{equation}\label{LagFunkt}
L:=T-V=-V\quad\textrm{for}\quad T=0.
\end{equation}

Since in MERW all paths that take the same amount of time are weighted equally, we have to take into account in MERW that the parameter time changes at different rates in different environments. The time evolution for a time step $\delta t$ is described by the formal Taylor series
\begin{equation}\label{ZeitEntw}
\psi(x,\tau+\delta t)=\exp\{\delta t\,\partial_\tau\}\psi(x,\tau).
\end{equation}

According to classical mechanics, the action $S$ generates the canonical transformations between initial coordinates and momenta and their time-dependent values. Using this relation, for the step matrix, we obtain
\begin{equation}\label{ZeitEntwOper}
\hat M\iist{ZeitEntw}{kSchritte}\exp\{\delta t\,\partial_\tau\}
=\exp\{\frac{S(\tau,\tau+\delta t)}{\hbar}\}
\ist{LagFunkt}\exp\{-\frac{V\delta t}{\hbar}\}
\ist{deltatdelta}\exp\{-\frac{V}{mc^2}\}.
\end{equation}

We work in an infinite cubic lattice in $R^D$ with lattice spacing $\delta$. The variation of the potential along the path in direction $\hat d$ is considered by the mid-point rule, well known from the path of integral formulation of quantum mechanics~\cite{Schulman1996}.
\vspace{-6pt}

\begin{equation}\label{MExp}
\hat M_d(\vec x):=\exp\{-\frac{V(\vec x+\hat d\delta/2)}{mc^2}\}\in\mathbbm R_+,
\quad d\in\{1,2,\dots,D\}.
\end{equation}

In every time step, determined by the Compton frequency, the particle has to move to a nearest neighbour. Under this assumption, the eigenvalue Equation~(\ref{EGlM}) for the D-dimensional problem reads
\vspace{-6pt}

\begin{equation}\begin{aligned}\label{EWGlPot}
\mathrm e^{\frac{V(\vec x)}{mc^2}}\,\lambda_i\,\psi_i(\vec x)&\ist{EGlM}
\mathrm e^{\frac{V(\vec x)}{mc^2}}\,\langle\vec x|M|i\rangle=\\
&\ist{MExp}\sum_{d=1}^D\left[\psi_i(\vec x+\hat d\delta)
\;\mathrm e^{\frac{V(\vec x)-V(\vec x+\hat d\delta/2)}{mc^2}}
+\psi_i(\vec x-\hat d\delta)
\;\mathrm e^{\frac{V(\vec x)-V(\vec x-\hat d\delta/2)}{mc^2}}\right].
\end{aligned}\end{equation}

To obtain potential differences on the right side of this equation, we multiplied \mbox{Equation~(\ref{EGlM})} by the factor $\mathrm e^{\frac{V(\vec x)}{mc^2}}$.

We expand Equation~(\ref{EWGlPot}) to increasing orders of $\delta$ and obtain in the zeroth order an equation for the approximate value of the eigenvalue
\begin{equation}\label{EWGlPot0}
\mathrm e^{\frac{V(\vec x)}{mc^2}}\,\lambda_i=2D+\mathcal O(\delta^2)
\end{equation}
which is not really solvable before knowing the effective potential for the eigenvalue $\lambda_i$. We will obtain this value at order $\delta^2$ only. We expand the expressions on the right side of Equation~(\ref{EWGlPot}) up to order $\delta^2$
\begin{equation}\begin{aligned}\label{EntwPsiExp}
\psi_i(&\vec x\pm\hat d\delta)\;
\mathrm e^{\frac{V(\vec x)-V(\vec x\pm\hat d\delta/2)}{mc^2}}=\\
&=[\psi_i(\vec x)\pm\delta\partial_d\psi_i+\frac{\delta^2}{2}\partial_d^2\psi_i]
[1\mp\frac{\delta}{2}\frac{\partial_dV}{mc^2}
-\frac{\delta^2}{8}\frac{\partial_d^2V}{mc^2}
+\frac{\delta^2}{8}\Big(\frac{\partial_dV}{mc^2}\Big)^2]+\mathcal O(\delta^3)
\end{aligned}\end{equation}
and get for Equation~(\ref{EWGlPot})
\begin{equation}\begin{aligned}\label{EWGlPotdeltaQ}
\mathrm e^{\frac{V(\vec x)}{mc^2}}\,\lambda_i\,\psi_i(\vec x)\ist{EWGlPot}
\sum_{d=1}^D&\left[\psi_i(\vec x+\hat d\delta)\;
\mathrm e^{\frac{V(\vec x)-V(\vec x+\hat d\delta/2)}{mc^2}}
+\psi_i(\vec x-\hat d\delta)\;
\mathrm e^{\frac{V(\vec x)-V(\vec x-\hat d\delta/2)}{mc^2}}\right]=\\
&\ist{EntwPsiExp}2D\psi_i+\delta^2\left[\Delta
-\frac{\vec\nabla V}{mc^2}\vec\nabla-\frac{1}{4}\frac{\Delta V}{mc^2}
+\frac{1}{4}\Big(\frac{\vec\nabla V}{mc^2}\Big)^2\right]\psi_i(\vec x).
\end{aligned}\end{equation}

In this equation, there appear two physical scales, the energy scale $mc^2$ and the length scale $\delta$. Their relation depends on the physical situation. For hydrogen, we expect the well-known relations of a dimensional analysis
\begin{equation}\label{HBohr}
E_N=-\frac{mc^2}{2}\frac{\alpha_f^2}{N^2},\quad
r_N=\frac{\hbar}{mc}\frac{N^2}{\alpha_f},\quad
V(r_N)=-\frac{\alpha_f\hbar c}{r_N}=-mc^2\frac{\alpha_f^2}{N^2}
\end{equation}
where $\alpha_f$ is Sommerfeld's fine structure constant. We conclude that the dimensionless ratios
\begin{equation}\begin{aligned}\label{Ordnungen}
\frac{V(r_N)}{mc^2}=\mathcal O(\alpha_f^2)\quad\textrm{and}\quad
\frac{\delta^2}{r_N^2}\iist{sigma1}{deltatdelta}
\frac{D(\frac{\hbar}{mc})^2}{r_N^2}=\mathcal O(\alpha_f^2)
\end{aligned}\end{equation}
are both of the order $\alpha_f^2$ allowing us to expand Equation~(\ref{EWGlPotdeltaQ}) to an increasing order in $\alpha_f^2$.

We reach order $\alpha_f^2$
\begin{equation}\label{VorSchGl2}
\lambda_i\,\psi_i(\vec x)\iist{EWGlPotdeltaQ}{deltatdelta}
(1-\frac{V}{mc^2})\,2D\,\psi_i+D\frac{\hbar^2}{m^2c^2}\,\Delta\psi_i.
\end{equation}

After multiplication with $mc^2/(2D)$, we obtain the stationary Schrödinger equation for a particle in a potential $V(\vec x)$
\begin{equation}\label{SchGl}
\big\{-\frac{\hbar^2}{2m}\Delta+V(\vec x)\big\}\psi_i(\vec x)
\ist{VorSchGl2}\underbrace{mc^2\Big(1-\frac{\lambda_i}{2D}\Big)}_{E_i}
\psi_i(\vec x)
\end{equation}
and can read off the relation between energy eigenvalues $E_i$ and eigenvalues $\lambda_i$ of the step matrix $M$. The higher the entropy, the lower the energy.

Expanding Equation~(\ref{EWGlPotdeltaQ}) up to $\alpha_f^4$, i.e., the exponential function up to the second order in $\frac{V}{mc^2}$, we obtain corrections to the Schrödinger equation
\begin{equation}\begin{aligned}\label{SchGlKorr}
\big\{-\frac{\hbar^2}{2m}\Delta+V(\vec x)
&\overbrace{-\frac{V^2(\vec x)}{2mc^2}+\frac{\hbar^2}{2m^2c^2}V(\vec x)\Delta
+\frac{\hbar^2}{2m^2c^2}\vec\nabla V\vec\nabla
+\frac{\hbar^2}{8m^2c^2}\Delta V(\vec x)}^{H_\mathrm{corr}}\big\}\psi_i(\vec x)=\\
&\iist{EWGlPotdeltaQ}{deltatdelta}\underbrace{mc^2\Big(1-\frac{\lambda_i}{2D}\Big)}_{E_i}
\psi_i(\vec x).
\end{aligned}\end{equation}

It is interesting to compare the correction terms $H_\mathrm{corr}$ with the correction terms derived from the Dirac equation~\cite{Thaller1992}
\begin{equation}\label{RelKorrKinEnergie}
H_{ck}=-\frac{\mathbf p^4}{8m^3c^2}=-\frac{T^2}{2m c^2},\quad
H_{\ell s}=\frac{1}{2m^2c^2}\frac{1}{r}\frac{\mathrm d V}{\mathrm dr}
\pmb\ell\mathbf s,\quad
H_D=\frac{\hbar^2}{8m^2c^2}\Delta V.
\end{equation}

The first term~$H_{ck}$ is the relativistic correction to the kinetic energy, the second term~$H_{\ell s}$ is the spin-orbit term. The third term $H_D$ is the Darwin term and usually attributed to Zitterbewegung of the electron.

We do not expect a relativistic correction to the kinetic energy in MERW, as it is a nonrelativistic model. Indeed, the corresponding first two terms in $H_\mathrm{corr}$ of Equation~(\ref{SchGlKorr}) cancel. This can be observed, using the virial theorem
\begin{equation}\label{Virial}
V=-2T,
\end{equation}
which is valid for the hydrogen atom already in the classical description. Further, using  the quantum mechanical operator for the kinetic energy
\begin{equation}\label{kinEne}
T=-\frac{\hbar^2}{2m}\Delta
\end{equation}
it follows
\begin{equation}\label{HckIst0}
\frac{V^2(\vec x)}{2mc^2}\ist{Virial}\frac{2T^2(\vec x)}{mc^2}
\iist{kinEne}{Virial}\frac{\hbar^2}{2m^2c^2}V(\vec x)\Delta.
\end{equation}

This shows that the relativistic correction to the kinetic energy vanishes in MERW, as it should. The spin-orbit term has some similarity to the third term in $H_\mathrm{corr}$, as both contain the gradient of the potential. But, as expected, they cannot really agree since MERW does not include spin degrees of freedom.

The last term of $H_\mathrm{corr}$ in Equation~(\ref{SchGlKorr}) has exactly the form of the Darwin term
\begin{equation}\label{GleichHcorrHD}
\frac{\hbar^2}{8m^2c^2}\Delta V(\vec x)=H_D.
\end{equation}

\begin{figure}[h!]
\includegraphics[scale=0.75]{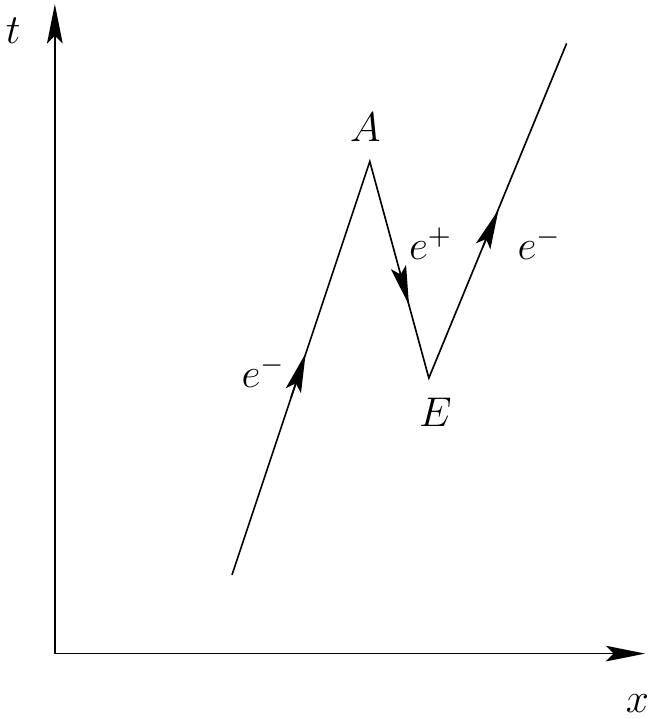}
\caption{Schematic picture of Zitterbewegung of electrons. An electron $e^-$ propagates in $x$-direction. Then, after some time, in $E$, an electron--positron pair is created. Electrons $e^-$ are represented by arrows in time direction and the positron $e^+$ by arrows backward in time $t$. At $A$, the original electron annihilates with the positron. By this process, the position of the original electron appears to be shifted by some accidental distance to a new position.}
\label{zitterbew}
\end{figure}

This term was first derived by Charles Galton Darwin when he solved the Dirac equation for the motion of an electron in a central force field~\cite{Darwin1928}. It can be derived in a general form in a nonrelativistic expansion of the Dirac equation~\cite{sakurai2006}. Usually, it is assumed that this term has no classical explanation~\cite{Chen2014}; it is interpreted as an electron--positron pair creation with the subsequent annihilation of the electron of hydrogen with the positron---see Figure~\ref{zitterbew}. In MERW, it does not need annihilation processes because it is a consequence of the diffusive motion. About the origin of this diffusive motion,one can only speculate.

\section{Excited States}\label{sec:Excited states}
As mentioned above, amplitudes of states with the highest entropy, quantum mechanical ground states, are non-negative and allow for an interpretation as numbers of trajectories. The case of a particle in a box with infinite walls shows that the increasing energy corresponds to increasing values of the negative Laplacian, see Equation~(\ref{SchGl}), and to an increasing number of nodes, where the amplitudes change their sign. Since amplitudes count numbers of trajectories, this poses a problem in the probability interpretation of trajectories in MERW. Analogous problems can already be identified in the first paper by Schrödinger 
on wave mechanics~\cite{Schrodinger:1926aaa} and in experiments on bouncing neutrons~\cite{Abele:2010a}. Schrödinger, after around two years of intensive deliberation (his notebooks from this time are available), decided to start from the Hamilton--Jacobi equation. He substituted the action $S$ by the exponential of $S$. Comparing the experiment, he realised that the natural unit of $S$ is the reduced Planck constant $\hbar$. Therefore, he equated $\psi= e^{S/\hbar}$. Observe that in this first paper he did not introduce the imaginary $\mathrm i$. There is a unique map of $S$ to $\psi$, but the inverse is not true. The nodes of $\psi$ and negative values of $\psi$ cannot be mapped to $S$. The existence of nodes also inhibits the particle interpretation of quantum mechanics, as one can nicely observe in the experiment with bouncing neutrons~\cite{Abele:2010a}. In this experiment, ultra cold neutrons are observed which are bouncing on a horizontal mirror with heights of roughly 30 \textmu m and a horizontal velocity of 6 m/s. The phase space calculations~~\cite{Suda2022} in this experiment clearly show that in an eigenstate of the corresponding Schrödinger equation, the neutrons cannot cross the nodes of the wave function. The general excuse is that the neutrons are never observed in an eigenstate, but always in a superposition. One has to conclude that true eigenstates are never realised in an experiment. On the other side, it is interesting to observe that in the momentum representation, one can clearly see how the neutrons fall with a vertical velocity, increasing proportionally in 
 time towards the reflecting mirror and bouncing back as expected with a constant deceleration; see Figure~20 of Ref.~\cite{Suda2022}.

Since amplitudes in MERW are normalised numbers of trajectories, they should not be negative. At first sight, negative values for amplitudes seem to contradict Kolmogorov's axioms of probability theory. However, there is an exit to avoid this contradiction. One can assume two species of particles with opposite signs of amplitudes. In the time evolution, both types of trajectories evolve towards the nodes and may cross the nodes. In numerical calculations, applying the step matrix $M$ according to Equation~(\ref{kSchritte}), one observes that the amplitudes of opposite signs annihilate each other. This nicely corresponds to trajectories of particles and antiparticles that can annihilate, as we observe them in high-energy physics experiments. With the particle--antiparticle interpretation, one can retain Kolmogorov's axioms and obtain a consistent probability interpretation of excited states as particle and antiparticle trajectories.

Of course it looks rather weird that we describe the spatial trajectories of bouncing neutrons in excited states by the oscillation between particle and antiparticle trajectories, whereas the momentum trajectories allow for a homogeneous interpretation. In a soliton interpretation of particles, this change in the interpretation may be related to the fact that one does not know what type of particles is present, as long as the particles do not react with the surroundings~\cite{Faber_2022}.

\section{Conclusions}
By its construction as a Markov chain with a step matrix $M$, Maximal Entropy Random Walk (MERW) determines the distribution of particle paths $\psi(\vec y,\tau)=\sum_{\vec x}\hat M^\tau(\vec y, \vec x)\,\mathring\psi(\vec x)$ starting from an initial distribution $\mathring\psi(\vec x)$ and paths of same duration $\tau$. The final distributions approach an equilibrium distribution, the time derivative term of the diffusion equation vanishes, and there is no difference between evolution forward and backward in time; thus, we arrive at the eigenvalue in Equation~(\ref{SchGlKorr}).

We observed many common properties between MERW and the time-independent Schrödinger equation:
\begin{enumerate}
\item  MERW treats stationary states of a diffusion process resulting from the eigenvalue Equation~(\ref{EGlM}) of the step matrix $M$. These states can be represented by rays in a Hilbert space.
\item MERW distinguishes between amplitudes $\langle0|x\rangle$ and probabilities $\rho_x$. The amplitudes count the number of trajectories arriving in the equilibrium distribution at a certain position. The probabilities indicate the share of trajectories, which, in the equilibrium distribution, pass at $x$. They are not modified by the Markov process. 
\item MERW respects the Born rule. It derives this rule from the relation~(\ref{pDef}) between the step matrix $M$ and the stochastic matrix $S$, and the request that $S$ leaves a probability distribution invariant. MERW impressively shows how the Born rule appears from a measurement between an equilibrium state in the past and the same equilibrium state in the future. This symmetry of the evolution from past to present, 
and  to the evolution from present to future, was already investigated by Erwin Schrödinger in his article~\cite{Schrodinger:1931aa} on the reversal of the laws of nature. The translation of this article is available in~\cite{Chetrite2021}.
\item With the appropriate choice of the diffusion constant~(\ref{DiffQM}), MERW derives the Schrödinger equation for a free particle.
\item Generalising the observation that time in a gravitational field is running in different heights with a different speed to arbitrary potentials, MERW respects the rule that all trajectories of same duration are counted with equal probability. In this way, MERW allows us to derive the Schrödinger equation for a particle in a potential.
\item From diffusion, MERW derives both the stationary Schrödinger equation and the Darwin term of the nonrelativistic expansion of the Dirac equation.
\end{enumerate}

If we want to use MERW to describe the properties of nature or to explain the background of the validity of the Schrödinger equation, we realise that MERW leaves several questions open or answers 
 them in the wrong way:
\begin{enumerate}
\item MERW is a diffusion process which propagates in $n^2$ the reduced Compton time $\bar t_c=\frac{\hbar}{mc^2}$ a distance of $n$ of reduced Compton wave lengths $\bar\lambda_c=\frac{\hbar}{mc}$. The width of the ground state of the atomic hydrogen may indicate $n=1/\alpha_f$ as a realistic order of magnitude. MERW does not explain the mechanism driving this diffusion process. The pair creation with a necessary energy of around 1~MeV may not be the origin of the Darwin term. After observing that in elastic scattering processes the position of the particles is not conserved, the author is rather inclined to think of some yet unknown scattering processes. In this respect, Ref.~\cite{EmergingQuantum2014} argues in favour of a zero-point field.
\item The kinetic term $-\frac{\hbar^2}{2m}\Delta$ in Equation~(\ref{SchGl}) is of a statistical origin only. The velocity seems to correspond to the diffusive velocity, as defined in~\cite{EmergingQuantum2014}~p.~125. In MERW, there is no ``classical'' kinetic energy, which is characteristic for the classical motion in Bohr's model of the hydrogen atom. In this sense, MERW cannot describe states with a non-vanishing angular momentum or oscillations in a classical potential. Such excited states appear in MERW as metastable states only, in distinction to the Schrödinger equation, where these states are stable. 
\item Amplitudes are numbers of trajectories and should be positive. This is true in the state with the highest entropy, corresponding to the ground state. Amplitudes for states with lower entropy have nodes and therefore regions with negative amplitudes. Trajectories traversing the nodes in opposite directions cancel and thus never cross the nodes. The particle--antiparticle picture has to be introduced to explain these cancellations.
\item MERW can explain the subtraction of trajectory numbers by a particle--antiparticle picture only. Therefore, the explanation of interference in Young's double slit experiment remains very questionable.
\item MERW does not explain the violation of Bell's theorem, as quantum mechanics nicely does.
\end{enumerate}

Due to these reasons, quantum mechanics cannot be simply a result of MERW, but MERW may provide some hints on how to proceed, in order to arrive at a better understanding of quantum mechanics and, like Mermin~\cite{Mermin1989}, not be satisfied with ``shut up and calculate``. With MERW, it is possible to derive the ground-state wave functions of particles in a potential and the exact form of the Darwin term. This could indicate that the ground state of quantum systems is determined by diffusive motion only.

Concerning the history of these ideas, we would like to mention that after a footnote of Eddington~\cite{Eddington:1927aa} concerning the probability in Schrödinger's wave mechanics, which is obtained by two symmetric systems of $\psi$-waves travelling in opposite directions in time, Schrödinger~\cite{,Schrodinger:1932aa,Schrodinger:1931aa} tried to find a probabilistic equation close to his wave equation. A probabilistic formulation of processes with initial and final densities was proposed by Bernstein~\cite{Bernstein1932} in 1932. Yasue suggested, in 1981, a time-symmetric variational approach with an action functional~\cite{Yasue1981}. In the formalism of stochastic processes, Zambrini studied Markovian processes of the Bernstein type in two papers and compared them to quantum mechanics~\cite{zambriniPhysRevA.33.1532,zambrini1986variational,zambrini1987euclidean}.

\acknowledgments{We acknowledge Open Access Funding by TU Wien. The author thanks to Jarek Duda for his explanations on Maximal Entropy Random Walk, further to  Jean Claude Zambrini for interesting discussions about Stochastic Processes and to John Peter Ralston about the interpretation of quantum mechanics.}

\appendix
\section[\appendixname~\thesection]{Entropy for Markov Processes}\label{SecEntropy}

In information theory, 1 Shannon, 
the unit of the information content is defined as 1~Bit. If in $N$ events the $i$-th type appears $N_i$ times, then its probability is
\begin{equation}\label{iTypWahr}
p_i:=\frac{N_i}{N},\quad N:=\sum_iN_i.
\end{equation}

With $n$ bits, we can distinguish between
\begin{equation}\label{AnzN}
N=:2^n
\end{equation}
events. If a low probability outcome really occurs, this has a high information content. The information content is therefore defined by
\begin{equation}\label{Infgehalt}
I(p_i):=\mathrm{lb}\frac{1}{p_i}=-\mathrm{lb}\,p_i
\end{equation}
and needs at least $I(p_i)$ bits to store it. Multiplying the information content $I_i$ of different outcomes with the number of their occurrences, we obtain the total number of bits
\begin{equation}\label{BitAnz}
n=\sum_iN_iI_i\ist{Infgehalt}-\sum_iN_i\mathrm{lb}\,p_i,
\end{equation}
necessary to store the information on the outcomes. The average number of bits defines the Shannon entropy $H$
\begin{equation}\label{iWahr}
H:=\frac{n}{N}\ist{BitAnz}-\sum_ip_i\,\mathrm{lb}\,p_i.
\end{equation}

The entropy of a process is the amount of information that is necessary to describe this process. A Markov process is defined by a probability distribution $\rho_x$ with the usual properties
\begin{equation}\label{DefProb}
0\le\rho_x\le1,\quad\sum_x\rho_x=1
\end{equation}
and by a transition matrix $s_{xy}$, which is a matrix of probabilities with
\begin{equation}\label{DefTrans}
0\le s_{xy}\le1,\quad\sum_ys_{xy}=1.
\end{equation}

\begin{figure}[h!]

\includegraphics[scale=0.8]{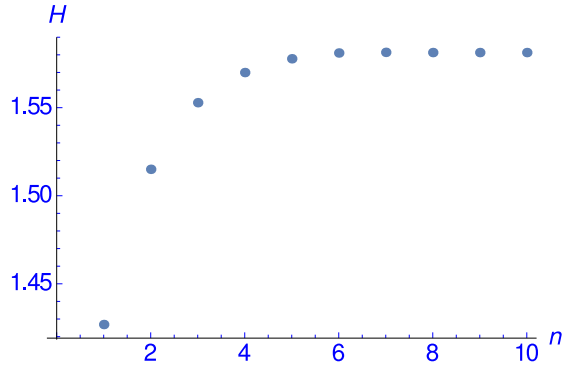}
\caption{Entropy $H$ for the process discussed in Figure~\ref{psirand} after $k=2^n$ steps.}\label{groundstateEntropy}
\end{figure}

A step beginning at $x$ leads with the probability $s_{xy}$ to $y$. According to Equation~(\ref{iWahr}), at least $-\mathrm{lb}\,s_{xy}$ bits are necessary to store the information about a step from $x$ to $y$. $k\rho_x$ of $k$ steps start at $x$ and $k\rho_xs_{xy}$ lead from $x$ to $y$. Therefore, at least $-k\sum_{x,y}\rho_xs_{xy}\,\mathrm{lb}s_{xy}$ bits are necessary, to store the information about this event. Thus, on average, the information content of a step is
\begin{equation}\label{EntropieZuw}
H=\sum_x\rho_x\sum_ys_{xy}\,\mathrm{lb}\frac{1}{s_{xy}}
\end{equation}
bits. Hence, $H$ is the average information production per step or in brief, the entropy. By definition, the transition probability $s_{xy}$ is the normalised number of paths leading from $x$ to $y$
\begin{equation}\label{SDefAllg}
s_{xy}=
\frac{\langle x|M|y\rangle\langle y|\psi\rangle}{\langle x|M|\psi\rangle}.
\end{equation}
$s_{xy}$ is undefined if the denominator is zero, i.e., no new path arrives at $x$. This has no influence on the value~(\ref{EntropieZuw}) of the entropy, since, for such sites, we have $\rho_x=0$. For all other sites $x$, $s_{xy}$ fulfills the requested conditions~(\ref{DefTrans}). Since $\psi$ changes from step to step, $s_{xy}$ is also modified in every step. For an infinite number of steps, $|\psi\rangle$ approaches the stationary solution $|0\rangle$ and $s_{xy}$ converges to $S_{xy}$ of Equation~(\ref{SInvar}). In Figure~\ref{groundstateEntropy}, we evaluate Equation~(\ref{EntropieZuw}) during the process shown in Figure~\ref{psirand}. As is clearly observed, $H$ increases during the process and approaches the entropy for the dominant eigenfunction $|0\rangle$ after an infinite number of steps.

For the stationary solution $|0\rangle$, we expect, for $k$ successive steps, an entropy production $H(k)=kH$. This expectation agrees with
\begin{equation}\begin{aligned}\label{lSchrittH}
&H(k):=\rho_x\sum_{yz}(S^k)_{xy}\,\mathrm{lb}\frac{1}{(S^k)_{xy}}\ist{SInvar}
\rho_x\sum_{yz}S_{xy}(S^{k-1})_{yz}\,\mathrm{lb}\frac{1}{S_{xy}(S^{k-1})_{yz}}
=\\&=-\sum_x\rho_x\sum_yS_{xy}\,\mathrm{lb}S_{xy}
\underbrace{\sum_z(S^{k-1})_{yz}}_{\ist{SIterNorm}1}-\sum_y
\underbrace{\sum_x\rho_xS_{xy}}_{\ist{SInvar}\rho_y}\sum_z(S^{k-1})_{yz}\,
\mathrm{lb}(S^{k-1})_{yz}=\\
&=-\sum_x\rho_x\sum_yS_{xy}\,\mathrm{lb}S_{xy}
-\sum_y\rho_y\sum_z(S^{k-1})_{yz}\,\mathrm{lb}(S^{k-1})_{yz}
\ist{EntropieZuw}H+H(k-1)=kH.
\end{aligned}\end{equation}

\begin{figure}[h!]
\centering 
\includegraphics[scale=0.8]{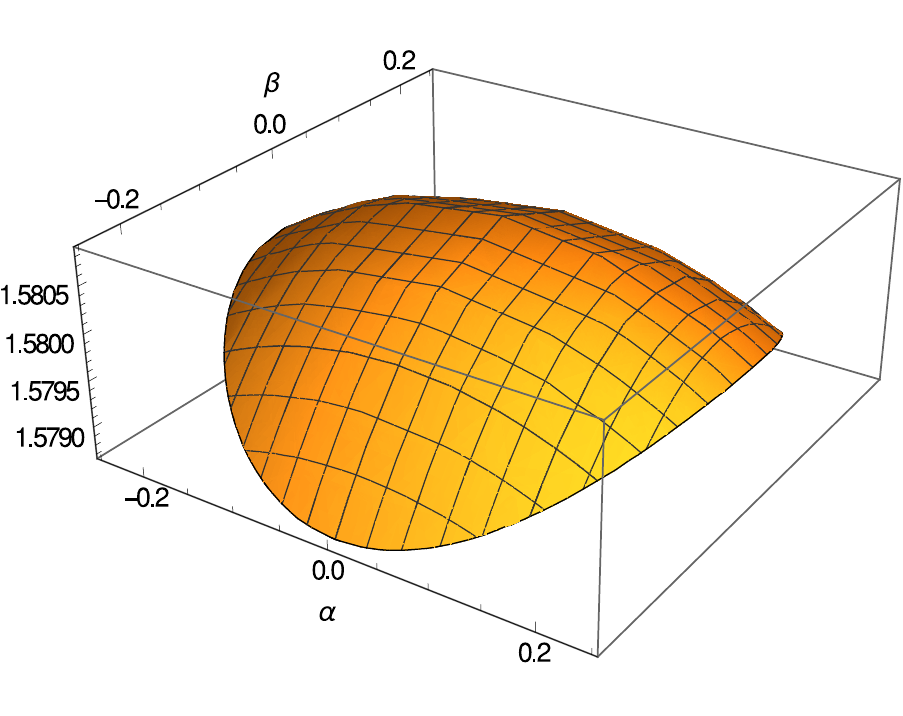}
\includegraphics[scale=0.8]{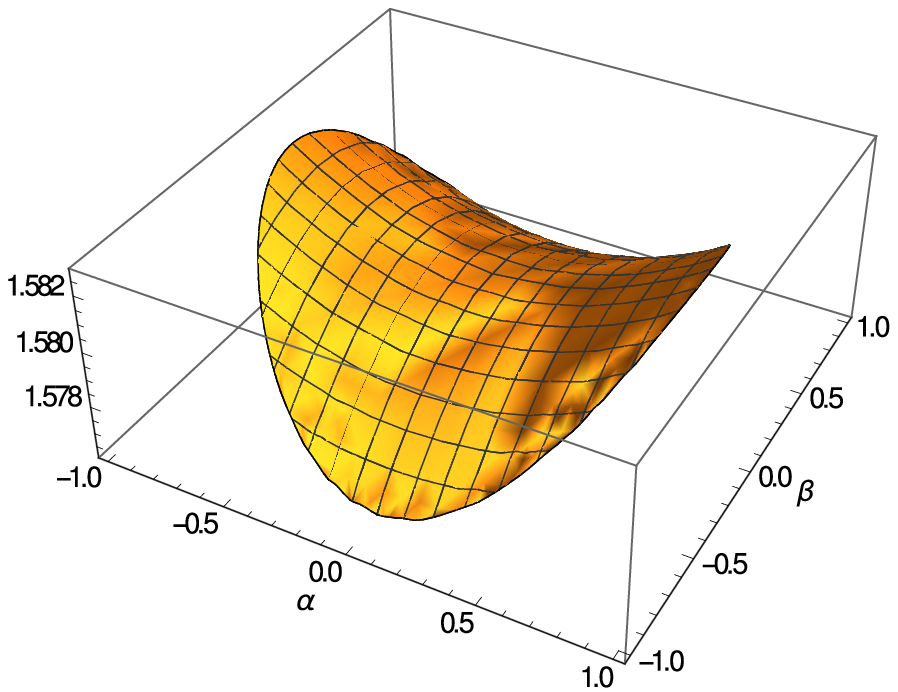}
\caption{The entropy production $H$ of Equation~(\ref{lSchrittH}) per step. In a Markov process, the states move towards higher entropy. Left: For $|\psi\rangle$ of Equation~(\ref{umpsi0}), we observe that the state $|0\rangle$ in the middle of the diagram has maximal entropy. The highest value on the boundary corresponds to an admixture of the state $|1\rangle$ and the lowest point corresponds to that of state $|2\rangle$. Right: For $|\psi\rangle$ of Equation~(\ref{umpsi1}), the state $|1\rangle$ is located at the saddle point in the middle of the diagram. The highest point of the boundary corresponds to the state $|0\rangle$ and the lowest point to the state $|2\rangle$.}
\label{psiEntrGE}
\end{figure}

According to Equation~(\ref{SchGl}), the ground state $|0\rangle$ of a particle in a potential $V$ belongs to the highest eigenvalue $\lambda_0$. As discussed in Section~\ref{SecMERW}, the eigenvector $|0\rangle$ of $M$ belonging to this highest eigenvalue $\lambda_0$ is also characterised by the largest entropy production $H$ per step of Equation~(\ref{EntropieZuw}). The value of $H$ for a free particle in a linear box $[-0.5,0.5]$ with 65~sites is depicted in the left diagram of Figure~\ref{psiEntrGE}, in the neighbourhood of the state $|0\rangle$ for
\begin{equation}\label{umpsi0}
|\psi\rangle=\sqrt{1-\alpha^2-\beta^2}|0\rangle+\alpha|1\rangle
+\beta|2\rangle,\quad\alpha^2+\beta^2\le0.06,
\end{equation}
where $|1\rangle$ and $|2\rangle$ are the first and the second excited state. As expected, the maximum of $H$ is at $\alpha=\beta=0$.

With increasing values of $\alpha$ and $\beta$, the mixed amplitude~(\ref{umpsi0}) can obtain nodes. This is the reason why, in the left diagram of Figure~\ref{psiEntrGE}, the values of $\alpha$ and $\beta$ are 
 restricted to $\alpha^2+\beta^2\le0.06$ and to values where no nodes appear. 
  This result suggests that we can determine the entropy of excited states from the stochastic matrix which results from blocking the nodes. With this method, we can show that excited states are saddle points of $H$. Admixing the ground state $|0\rangle$ to the first excited state $|1\rangle$ increases $H$. All other admixtures decrease $H$. In the right diagram of Figure~\ref{psiEntrGE}, the second excited state $|2\rangle$ contributes to $|\psi\rangle$
\begin{equation}\label{umpsi1}
|\psi\rangle=\alpha|0\rangle+\sqrt{1-\alpha^2+\beta^2}|1\rangle
+\beta|2\rangle,\quad\alpha^2+\beta^2\le1.
\end{equation}

The lowest point at the boundary of the saddle-shaped surface corresponds to an admixture of the ground state and the highest point corresponds to an admixture of the state $|2\rangle$.

As excited states are metastable, a small disturbance is sufficient to let them move towards the ground state,\ which has the highest entropy.

\bibliography{references}

\end{document}